\def\c4{C\,{\sc iv}}
\def\n5{N\,{\sc v}}
\def\o6{O\,{\sc vi}}
\documentstyle[11pt,paspconf]{article}

\begin{document}

\title{A Review of Line and Continuum Correlations in AGNs}
\author{Patrick S. Osmer}
\affil{Department of Astronomy, The Ohio State University,
174 W. 18th Ave., Columbus, OH 43210, posmer@astronomy.ohio-state.edu}
\author{Joseph C. Shields}
\affil{Department of Physics \& Astronomy, Ohio University,
Athens, OH 45701, shields@phy.ohiou.edu}

\begin{abstract}
We review the observational evidence for the Baldwin Effect, and the
empirical behavior of emission lines in relation to multi-wavelength
continuum emission for AGNs.
\end{abstract}

\section{Introduction}

It is a pleasure to review the observational basis for the Baldwin
Effect (BE).  Our meeting is timely for several reasons: the
considerable progress in the field since the discovery of the BE,
improved theoretical understanding of the emission-line region in
quasars and active galactic nuclei (AGNs), the significant advances in
multi-wavelength observations (from X-rays through to radio), and the
prospects for future work that the new 8-10-m ground-based telescopes
and space observatories will enable.

It is also appropriate that Jack Baldwin and Gary Ferland have taken
the lead in organizing the meeting: Jack for his discovery and subsequent
work on the BE and Gary for his work on developing theoretical
and computing tools for understanding the emission-line regions
of quasars.

We are glad to see Joe Wampler here.  His work with Lloyd Robinson on 
developing the image tube scanner for the Lick 3-m telescope enabled
Jack to make the observations that led to the discovery of the BE.
The scanner was the first instrument that offered the requisite sensitivity,
linearity, and broad wavelength coverage needed for the systematic study of 
emission lines in quasar spectra.

This paper is organized as follows.  We give a brief history of the
BE in the next section and a discussion of the selection and other
effects that are important to its determination and interpretation.
That is followed in \S3 by a review of the current status
of the BE, including different observational methods and analyses and
the presence of the BE in other emission lines besides \c4.  
In \S4 we consider more detailed luminosity effects, and in
\S5 we review the intrinsic BE, i.e., how the equivalent
widths of emission lines change when the continuum level in a particular
object varies.  In \S6 we discuss further systematics and phenomenology,
such as which parts of the line profiles exhibit the BE and the X-ray BE.
Then in \S7 we review the BE in high-redshift objects and conclude in
\S8 with a summary and topics for future work. 

\section{History}

\subsection{Discovery}

\begin{figure}
\vspace{3in}
\includegraphics{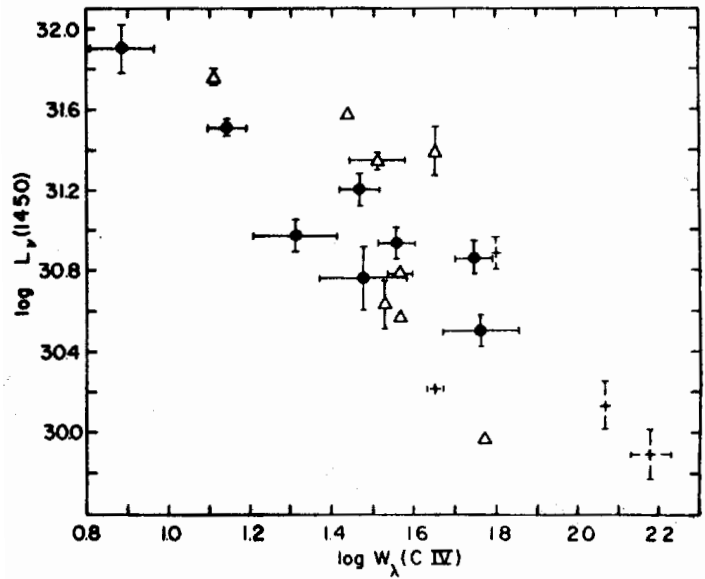}
\caption{Baldwin's (1977a) plot of the BE for \c4.}
\label{figOA}
\end{figure}

The discovery paper for the BE was Baldwin's (1977a) article, which
noted an anti-correlation of the equivalent width $W_\lambda$ of the \c4
emission line and the continuum luminosity $L$ of quasars on
the assumption that their redshifts were cosmological.  A copy of his
original diagram is shown in Figure~\ref{figOA}.  The effect was
pronounced: the equivalent width of the lines decreased by an order of
magnitude as the continuum luminosity increased by 1.6 dex.

Baldwin (1977b) published another seminal paper that same year in
which he noted that the observed average Ly$\alpha$/H$\beta$ intensity
ratio in quasar spectra was significantly less than predicted for a
recombination spectrum.  Subsequently Baldwin et al. (1978) showed
that the Mg II $\lambda$2798 emission line also exhibited an
anti-correlation with luminosity and used the BE to constrain
cosmological models.  They concluded that their data ruled out the
local hypothesis for quasars and zero-pressure models with $q_{0} =
0$.

The first two papers mentioned above stimulated much work on quasar
emission lines in the subsequent two decades.  The third one pointed
out the cosmological possibilities of the BE; however, the
uncertainties were large enough to limit the usefulness of the
technique, a problem that has continued to this day.  One of the goals
of this meeting is to see if our knowledge has advanced enough for
quasars to be useful cosmological probes.

\subsection{Doubts and Confirmation}

After the original BE discovery paper (Baldwin 1977a), doubts were
raised about its nature, reality, or utility for cosmological studies.
Although we now regard the BE as well established, there are several
effects that must be considered in any analysis of the BE:

\begin{enumerate}

\item Selection Effects.  Jones \& Jones (1980) pointed out that faint 
quasars with weak emission lines would be difficult to observe and
therefore cause a selection effect that could artificially enhance the
apparent BE.  Osmer (1980) showed that the slitless spectrum technique
would systematically favor the discovery of strong-lined quasars at
faint magnitudes in a manner that was also in accord with the apparent
BE.  More recently, Yuan et al. (1998) have shown that different
dispersions in observational parameters can artificially produce
luminosity correlations in data sets.

\item Variability.  Murdoch (1983) discussed how the variability of
flat-spectrum quasars could account for much of the BE.  For example,
if the line emission stayed constant while the continuum decreased in
brightness, the equivalent width would appear large when the object
was fainter.  In reality, both the line and continuum vary with time,
which makes the situation more complicated.  In any case, the effect
of variability must be considered in analyses of the BE.

\item Population Effects.  Baldwin himself (1977a) noted in the BE discovery
paper that the flat-radio-spectrum quasars showed the tightest
correlation in the log\,$L_\nu$(1450\AA) $-$ log $W_\lambda$(\c4)
diagram.  A long-standing question has been whether other
types of quasars exhibit the same BE as do the flat-spectrum objects.
This question bears on the physical nature of the different types of
quasars and AGNs as well as their usefulness as cosmological
indicators.

\item Evolution.  If the physical nature of quasars varies with cosmic time,
then the BE observed at one redshift might not be appropriate for
quasars at a different redshift. The separation of evolutionary
(redshift) effects from luminosity effects has been difficult in the
flux-limited samples studied to date.  The steep rise of quasar number
counts with magnitude causes most of them to appear near the magnitude
limit.  This in turn will introduce a correlation of redshift and
intrinsic luminosity into the data set in that the most luminous
quasars will be those with the largest redshifts.

\end{enumerate}

Now we will consider how all these effects have played out in subsequent 
studies of the BE.  

Baldwin, Wampler \& Gaskell (1989, BWG), building on earlier work,
addressed the problem of selection effects for radio-loud quasars by
obtaining spectra of all objects in a well-defined radio sample, so
that objects with weak lines would not be omitted.  For optically
selected quasars, they used quasars selected by the ultraviolet excess
(UVX) technique, which was based on quasars (primarily those with $z <
2.2$) being much brighter in the ultraviolet than most stars and was
less dependent on emission-line properties than the slitless spectrum
technique\footnote{All observational techniques have selection effects
(see, e.g., Wampler \& Ponz [1985] and Peterson [1997]).  The
important thing is to account for them properly, which can now be done
with well-defined samples based on quantitative selection
techniques.}.  BWG demonstrated that the BE was indeed present in
radio-loud quasars.  They also pointed out that their data could not
establish differences between the BE in the radio- and UVX-selected
quasars because of the differences in the luminosity limits of the
samples.

\begin{figure}
\vspace{6.5in}
\includegraphics{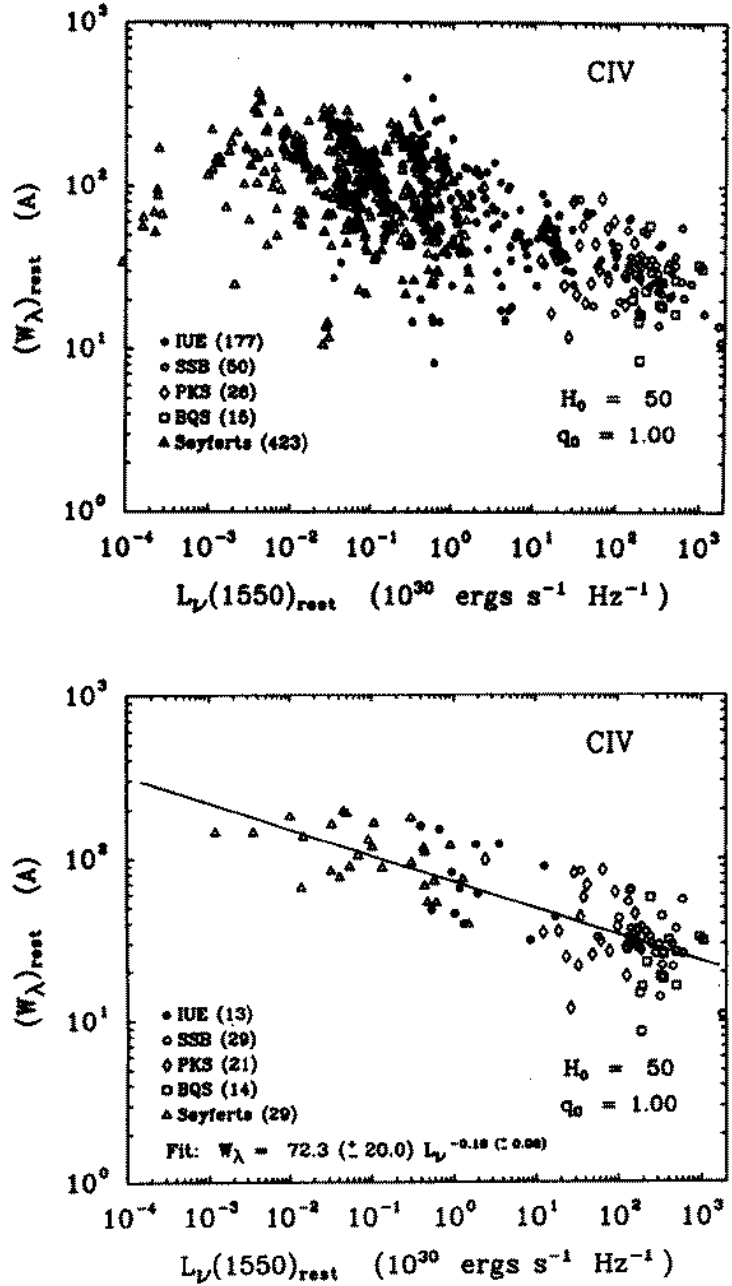}
\caption{The Kinney et al. (1990) results for the \c4 BE.
The upper panel shows results for the full {\sl IUE} data set,
supplemented by ground-based measurements by Sargent et al. (1989) and
by BWG.  The lower panel shows the same plot for a subset of data
selected for high signal-to-noise ratio, with repeated observations
averaged into a single point per object.
}
\label{figOB}
\end{figure}

A next important step in confirming the reality of the BE was made by
Kinney, Rivolo, \& Koratkar (1990, KRK), who used {\sl IUE} archival data
for quasars and Seyfert galaxies to obtain a range of $10^7$ in
continuum luminosity.  Their data showed conclusive evidence for the
BE being a real, physical effect (Fig.~\ref{figOB}, upper).
Furthermore, they were able to address the variability problem and
show that it caused much of the scatter in their overall set of data.
When they averaged multiple observations of different objects into
single points, the scatter was much reduced (Fig.~\ref{figOB}, lower).

Thus, by 1990, there was convincing evidence for the reality of the BE
as well as for the importance of selection effects and variability,
which is now known as the intrinsic BE.  The intrinsic BE is an
important subject in itself, and we discuss it in more detail below in
\S 5.

The question of population effects on the BE has been more difficult
to settle, with different investigators finding different results.
One problem is that the intrinsic scatter in many of the properties of
quasar and AGN spectra can mask effects like the BE, especially in
small samples or ones that do not cover a sufficiently large range in
luminosity, as the work of BWG and KRK made clear.  A further
complication when comparing radio-loud and radio-quiet systems is the
necessity of working with samples that span the same range in
luminosity, a point emphasized by BWG.  Sargent et al. (1989) and
Steidel \& Sargent (1991) compared the behavior of radio-loud and
radio-quiet sources matched in luminosity, and found indications
supporting Baldwin's claim of a stronger BE correlation for radio-loud
QSOs\footnote{The radio-loud objects studied by Steidel \& Sargent
(1991) were primarily flat-spectrum (core-dominated) systems, although
they found no significant differences between the behavior of flat-
and steep-spectrum quasars within their sample.}.  However, a visual
scrutiny of their results suggests that this is a question that would
benefit from the study of larger samples spanning a broader luminosity
range; the significance of correlations in the published plots often
rests on the location of a very small number of points.

Zamorani et al. (1992) combined all data sets available at that time
for optically selected quasars to make an improved investigation of
the possible differences between radio- and optically-selected
quasars.  From data on the \c4 line in 316 quasars, they found the
presence of a BE, but with a slope about half the value of BWG for
radio-selected quasars.  Their slope was in good agreement with the
KRK results, but the normalization was about half as large.  They
considered that variability of the quasars in the BWG sample with the
strongest CIV emission was an important factor causing the difference
between the radio and optically selected quasars.  In the end,
however, they did not think it possible to conclude there was a
physical difference between the two classes of objects.  A difficulty
with this type of analysis is the inherent inhomogeneity in the data when
equivalent widths are compiled from the literature; a large scatter in
observed equivalent widths can result artificially from the disparate
measurement methods employed by different researchers.  Regrettably,
we still appear to be in the situation where a comparison based on
even larger and more carefully selected samples is needed.

Finally, the question of evolutionary effects in quasar spectra
continues to be important and is just now becoming feasible to
address.  Here the difficulty has been in obtaining samples and
especially follow-up slit spectroscopy of faint quasars/AGNs at high
redshift.  For samples with limiting magnitudes of, say, magnitude 21,
objects at $z = 3$ will have $M < -25\ (H_0=75,\ q_0=0.5)$, far more
luminous than nearby AGNs.  A key project for large ground-based
telescopes will be the discovery and spectroscopy of quasars with $z
\geq 3$ and $m \geq 25$ --- with such data a systematic and
significant investigation of the evolutionary properties of quasar
spectra will become possible.

\section{Current Status}

\subsection{Observational Methods and Results}

Our understanding of the BE has grown in recent years through application of
several distinct approaches to the study of AGN emission lines.  These
techniques include:

1) Measurement of lines in high signal-to-noise ratio spectra, with
bivariate analysis.  This approach is the simplest to grasp, and the
same as that employed by Baldwin (1977a), but improvements in
technology have made it possible to study ensembles with very high
signal-to-noise ratio.  This method arguably reached its pinnacle in
the studies by Laor et al. (1994, 1995) of quasar spectra acquired
with {\sl HST}.  The data were of sufficient signal-to-noise ratio to permit
deconvolution of line blends and accurate measurement of line wings,
which are often lost in spectra with only modest signal-to-noise
ratio.  With a sample of only 18 QSOs, Laor et al. were able to
discern a BE in nearly all of the lines of at least moderate strength
between 1000 $-$ 2000 \AA\ (9 features in total), with the notable
exception of \n5 $\lambda$1240.

\begin{figure}
\vspace{3.3in}
\includegraphics{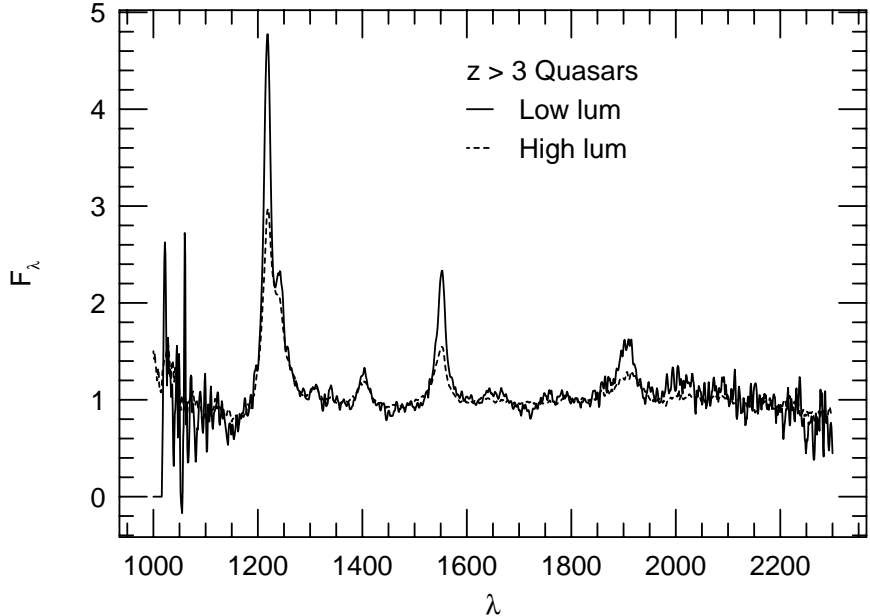}
\caption{The OPG composite spectra for high- and low-luminosity quasars
with $z > 3$.}
\label{figOD}
\end{figure}

2) Construction of composite spectra.  Subdividing a sample into
luminosity intervals, and subsequently generating an average or median
spectrum for each bin, offers several advantages for identifying
luminosity-dependent behavior.  Combining spectra into a composite
decreases the shot noise present in any individual spectrum, and also
diminishes the ``noise'' contributed by the intrinsic peculiarities of
any single source.  Composites permit identification of
luminosity-dependent line effects by simple inspection, or more
quantitative methods.  Working with actual spectra, rather than line
measurements, also makes it possible to distinguish
luminosity-dependent behavior across the profiles of individual lines.
Examples of this approach include studies by V\'eron-Cetty et
al. (1983), Osmer et al. (1994, OPG), Laor et al. (1995), and Francis
\& Koratkar (1995).  An example of such a comparison is shown in
Figure~\ref{figOD}, which compares composite spectra assembled by OPG
for high- and low-luminosity quasars at $z > 3$.  The BE is apparent
from inspection of most of the stronger lines, and the direct
comparison also reveals qualitative differences in the strength of the
effect (e.g. large variation in \c4 but only a small difference in the
1400 \AA\ feature).

\begin{figure}
\vspace{6.2in}
\includegraphics{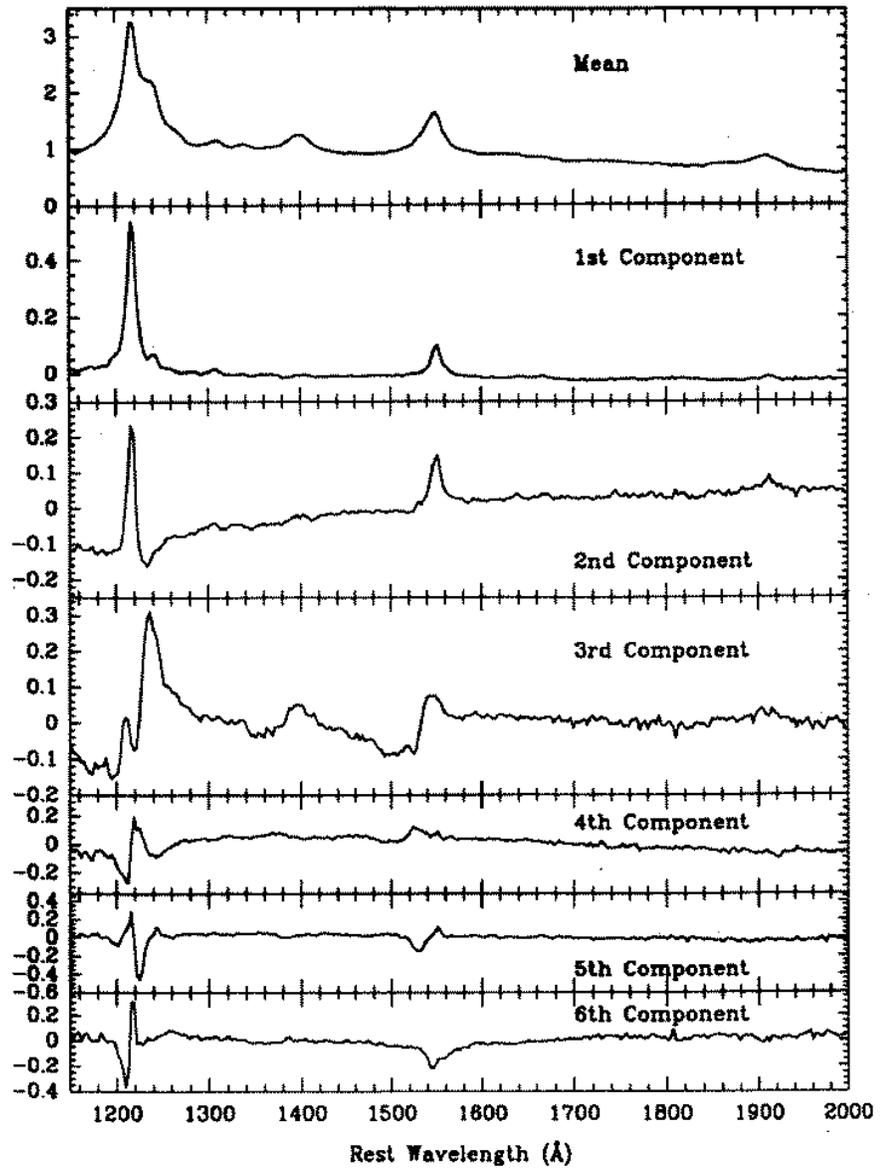}
\caption{The Francis et al. (1992) principal components found for the LBQS.}
\label{figOC}
\end{figure}

3) Principle component analysis (PCA).  PCA examines correlations
between the variances of parameters measured for a sample, and builds
a minimum set of basis vectors across this parameter space that
describes the total variance in the system (see Wills \& Francis, this
proceedings).  PCA has been employed in two modes in the study of
quasar spectra.  The first directly examines variations in spectra,
with the measured parameters comprised of the flux density as a
function of wavelength after putting the spectra on a common
normalization.  Francis et al. (1992, FHFC) used this method to study
the spectra of 232 objects with $1.8 < z < 2.7$ from the Large Bright
Quasar Survey to classify the ultraviolet features.  They found that
the first three principal components accounted for about 75\% of the
observed variance in the spectra (Fig.~\ref{figOC}).  These components
can be interpreted approximately as largely independent variables
describing the strength of the emission line cores, the slope of the
continuum, and the prominence of broad absorption line features.

The alternative application of PCA operates on a parameter set of
quantities measured from the spectra, such as line equivalent widths,
profile indices, etc.  Boroson \& Green (1992) published an
influential study of this type, using optical rest-frame spectra of 87
QSOs in the Bright Quasar Survey with $z < 0.5$.  In this case,
61\% of the observed variance in the spectra was accounted for with
three principal components, and the first eigenvector is dominated by
an anticorrelation between measures of Fe~{\sc ii} and [O~{\sc iii}]
line strength.

Conventional studies of line strengths as well as the innovations
discussed above have led to reports of a BE in lines of Ly$\beta$
$\lambda$1026, O~{\sc vi} $\lambda$1034, Ly$\alpha$ $\lambda$1216,
O~{\sc i} $\lambda$1304, C~{\sc ii} $\lambda$1335, \c4 $\lambda$1549,
He~{\sc ii} $\lambda$1640, Al~{\sc iii} $\lambda$1857, C~{\sc iii}]
$\lambda$1909, and Mg~{\sc ii} $\lambda$2798.  Given the length of this
list, it is of interest to examine those lines that apparently do {\sl
not} participate in the BE.  Most studies have shown little or no
evidence of a BE for the Si~{\sc iv}+O~{\sc iv}] $\lambda$1400 blend
(e.g., Cristiani \& Vio 1990; FHFC; OPG; Francis \& Koratkar 1995).
The interpretation of this result is unclear, and the situation
remains ambiguous in light of a strong BE signal for this feature
reported by Laor et al. (1995).

Stronger concurrence is available for the behavior of \n5, which
essentially all studies indicate does not show a BE; the equivalent
width of the line is apparently independent of source luminosity
(e.g., OPG; Francis \& Koratkar 1995; Laor et al. 1995).  Hamann \&
Ferland (1992, 1993) have argued that this behavior can be understood
as an abundance effect, with \n5 emission enhanced relative to other
lines in luminous sources resulting from selective (secondary)
enrichment of nitrogen in vigorously star-forming environments.
Expressed in terms of line intensity ratios, the \n5/\c4 and
\n5/He~{\sc ii} ratios are enhanced in more luminous systems, and the
study by OPG suggests that this is truly a luminosity, rather than
redshift, effect.  Under this interpretation of the \n5 line strength,
the decoupling of $W_\lambda$(\n5) and $L$ would result from the
luminosity dependence of N enrichment largely canceling out the normal
BE signature.

In the optical bandpass, several studies have addressed the luminosity
dependence of broad H$\beta$, which also presents little evidence of a
BE (e.g., Yee 1980; Binette et al. 1993; Boroson \& Green 1992).  Only
very limited information has been published on the luminosity
dependence of the narrow emission lines.  Baldwin (1987) has argued
that the ultraviolet lines in luminous QSOs are unexpectedly weak, in
comparison with Seyfert 2 galaxies (see also Wills et al. 1993b; Laor
et al. 1994, 1995).  Boroson \& Green (1992) found a positive
correlation, significant at $>$99\% confidence, between absolute
magnitude $M_V$ and $W_\lambda$([O~{\sc iii}] $\lambda$5007) for PG
QSOs, consistent with a BE.  Brotherton (1996) performed a PCA
analysis on radio-loud quasars, and while no BE was apparent in a
simple bivariate analysis for this sample, his first eigenvector
displayed strong BE-like behavior in H$\beta$ and [O~{\sc iii}].  This
trend is evidently masked by scatter from other parameters in the full
spectra, leading to the lack of a simple correlation.

Even if the narrow lines do exhibit a BE, the extent to which this
behavior shares an origin with the broad-line BE may be open to
question.  Wills et al. (1993b) have argued that the weakness of the UV
narrow lines in luminous sources may be due to reddening by dust, part of
which is associated with the narrow-line region (NLR) itself; but the
broad-line region (BLR) is unlikely to harbor significant dust (e.g.,
Laor \& Draine 1993).  More generally, the NLR in QSOs may extend to
kpc scales, at which point the properties of the emitting gas may be
dictated primarily by the distribution of matter in the host galaxy
rather than more exotic processes operative on small scales.  If the
NLR plasma resides in a disk with a scale height $h$ independent of
the luminosity $L$, a BE would in fact be expected.  The length scale
$r$ characteristic of the NLR will move out (increase) with $L$.  The
solid angle describing NLR coverage scales with $h/r$, so that the NLR
covering factor, and hence equivalent width $W_\lambda$ of line
emission, will decrease in more luminous systems.  This scenario is a
variant of the ``receding torus'' model described by Lawrence (1991;
see also Simpson 1998), whereby a dusty torus obscures less of the
central nucleus in more luminous systems; the dust sublimation radius
grows with increasing $L$.  In this picture, NLR emission arises in
the unobscured cones along the torus axis.  As the torus recedes, the
cones expand in opening angle, and the covering factor of NLR gas may
then {\sl increase} with larger $L$.  The luminosity dependence of
narrow-line emission in AGNs may thus yield interesting insights into
AGN structure, but may have only limited connection to the classical
broad-line BE\footnote{The ambiguity inherent in ascribing common
physical processes to sources with similar phenomenology is
underscored by the claim that Wolf-Rayet stars exhibit a BE (Morris et
al. 1993).}.

\subsection{Slope as a Function of Ionization}

For those lines that show a BE, the growth in observations now makes
it possible to go beyond testing for correlations, by measuring and
comparing the slopes of $W_\lambda$ versus $L$ for different lines.
Several studies have completed this exercise, and produced evidence
that lines of relatively high ionization, such as \c4, exhibit
systematically steeper slopes in the Baldwin diagram than lines of
lower ionization, such as Mg~{\sc ii} or Ly$\alpha$ (e.g., Wu et al.
1983; Kinney et al. 1987; KRK; Baldwin et al. 1989; Zheng et
al. 1997).  This result is sometimes expressed in terms of the
luminosity behavior of line ratios, such that, for example,
Ly$\alpha$/\c4 and C~{\sc iii}]/\c4 appear to increase systematically
with increasing luminosity.  The existence of this trend was confirmed
at this meeting by Brian Espey, who reported results for the BE based
on a very large sample of AGN spectra.  Zheng, Kriss, \& Davidsen
(ZKD, 1995) found a particularly steep BE for the O~{\sc vi} line,
consistent with its high ionization.  ZKD's study benefited from
inclusion of {\sl HUT} spectra for low-redshift objects, which
resulted in a wide baseline in luminosity for analysis.  An apparently
contradictory finding was reported by Paul Green (1996), however, who
found {\sl no} statistically significant BE in O~{\sc vi} for an
overlapping sample drawn from the {\sl IUE} archive; Green emphasized
the importance of including upper limits in the Baldwin diagram when
testing for correlations.  Nonetheless, Green considered the ZKD
result to be valid because of their detection of \o6 in all their
objects and because their sample covered a larger range in luminosity
than his\footnote{A further potential complication for study of the
\o6 BE is the influence of the unresolved Ly$\alpha$ forest.  In many
samples, luminosity is strongly correlated with redshift, so that
high-luminosity objects are at relatively high $z$; features shortward
of the Ly$\alpha$ emission line may thus be preferentially corrupted
in the high-luminosity systems.  Since the Ly$\alpha$ forest affects
both the \o6 line and adjacent continuum, in a formal sense this is
unlikely to produce an artificially steeper BE, but should merely
increase the scatter at high luminosity.  But a bias may nonetheless
result if line measurement algorithms are not robust to the diminished
signal-to-noise ratio and modified line profiles resulting from
Ly$\alpha$ forest absorption.  Future studies of the
\o6 feature would benefit from some simple modeling of the statistical
influence of the Ly$\alpha$ forest on measured line strengths.}.

The fact that different lines show different slopes in the Baldwin
diagram is important for understanding the BE theoretically.  A simple
luminosity dependence of broad-line-region (BLR) covering factor for
the central continuum source should lead to similar variations in
$W_\lambda$ for all lines.  Likewise, simple models invoking
inclination variations of accretion disks (which produce an
anisotropic continuum radiation field) as the source of the BE do not
predict different slopes for different lines.  The trend of steeper
slopes for lines of higher ionization is an important clue that must
be accommodated by alternative explanations for the BE.  Dust internal
to the BLR could potentially introduce BE-like behavior if cloud
properties conspire to produce more absorption for higher $L$ (Shuder
\& Macalpine 1979), but is unlikely to produce the observed
ordering, even if dust were able to survive in this inhospitable
environment.

Several scenarios that take into account the ionization dependence of
the BE have been advanced.  Mushotzky \& Ferland (1984) argued that
this behavior was consistent with a luminosity dependence of the
ionization parameter $U$ (ratio of ionizing photon and particle
densities) for the BLR clouds, such that $U$ was less in
more luminous sources.  Since $U$ provides an indicator of the degree
of ionization for a photoionized cloud, this prescription takes
explicit account of the BE ionization dependence, although the
underlying cause of a $L - U$ correlation is not obvious.  This model
does not predict a BE for Ly$\alpha$, however, and this line might actually
be expected to {\sl increase} in equivalent width as $U$ diminishes
(Shields \& Ferland 1993); an additional luminosity dependence of
covering factor may be required to fit the observations.  Shields et
al. (1995) recast this formulation as a luminosity dependence of
coverage by high-$U$ clouds; this scenario was motivated in part by
reverberation-mapping evidence for matter-bounded clouds, coupled with
a luminosity dependence in outflow (signaled by narrow blueshifted
resonance absorption in Seyferts versus broad absorption lines in
QSOs) that may affect the coverage of tenuous components of the BLR.

A final interpretation that is gaining considerable support views the
ionization trend in the BE as the consequence of a
luminosity-dependent continuum spectral energy distribution (e.g.,
Schulz 1992; Netzer et al. 1992; Zheng et al. 1992; Zheng \& Malkan
1993; Green 1996, 1998; Wang et al. 1998; Korista et al. 1998).  In
this picture, more luminous AGNs feature softer ionizing continua; the
reduction in ionizing photons at a given optical/UV luminosity leads
to smaller equivalent widths, with the high-ionization lines
responding most strongly, due to their greater sensitivity to the
high-energy continuum.  This explanation has several attractive
features.  First, some accretion disk models naturally predict a
softer continuum for more luminous systems (e.g., Netzer et al. 1992).
Second, observational evidence exists in direct support of the
requisite continuum behavior.  These data are often cast in terms of the
luminosity dependence of $\alpha_{ox}$, the two-point spectral index
connecting 2500 \AA\ and 2 keV in the source rest-frame (e.g.,
Zamorani et al. 1981).  The role of selection effects and
photometric errors in defining these trends continues to be of
concern, however (e.g., La Franca et al. 1995), and a recent paper
by Yuan et al. (1998) on this subject provoked much discussion at this
conference.

\section{Detailed Luminosity Effects}

\subsection{The Importance of Luminosity Range}

In \S2.2 we commented on the importance of having a sufficiently
large range in luminosity to detect and study the BE.  Selection
effects and intrinsic scatter in the properties of quasar spectra
can mask the appearance of the BE.  We also mentioned that the KRK
study provided some of the most conclusive evidence for the BE because
their sample spanned $\sim$ 7 orders of magnitude in luminosity. 

The fact that the BE is seen over such a large luminosity range has
theoretical implications, in addition to observational uses.  This
finding provides another challenge to models that rely solely on disk
inclination effects or other sources of continuum beaming to produce a
BE (KRK; Francis 1993); these scenarios can generally reproduce a
BE-like correlation over only $1 - 2$ orders of magnitude in $L$.
Variability was suggested early on as the source of the BE (Murdoch 1983),
but the typical amplitude of variability seen to date in Seyfert nuclei
and normal QSOs is $\la 1$ order of magnitude, rather than 7, and
when AGNs {\sl do} vary, they appear to trace out a locus in the
$W_\lambda - L$ plan that is distinct from that of the ensemble BE
(KRK; see also \S5).  

Given the existence of large scatter in the Baldwin relations, is it
possible to understand the origins of this dispersion?  Inclination or
other beaming phenomena, if not responsible for the BE, may provide a
source of scatter in the correlation (Netzer et al. 1992); however,
this effect is unlikely to dominate the scatter, since predictions of
the resulting equivalent width distributions appear to be inconsistent
with the observations (Francis 1993).  Variability clearly {\sl is} a
source of scatter in the BE, at least at low luminosities.  Korista et
al. (1998) have recently suggested that metallicity is another
probable source of scatter in the overall trend.  This last hypothesis
may be testable via use of the N~{\sc v} line strength as a crude
indicator of abundances within the broad-line media.

\subsection{Second Order Effects}

While the BE is usually parametrized quantitatively by the slope of a
straight line in the $\log W_\lambda - \log L$ diagram, there is considerable
evidence suggesting that curvature exists within the Baldwin trends.
The sense of this behavior is that the BE exhibits a steeper slope at
higher luminosities (e.g., V\'eron-Cetty et al. 1983; Wu et al. 1983;
KRK; OPG).  The flattening at low luminosities has resulted in
suggestions by some authors that Seyfert nuclei do not participate in
the BE, but instead have equivalent widths independent of luminosity
(e.g., Wampler et al. 1984).  Discerning luminosity-dependent behavior
within Seyfert ensembles is again complicated by substantial scatter
about any underlying trend, with much of the dispersion stemming from
intrinsic variability (e.g., KRK).  What seems clear from existing
studies, however, is that Seyferts connect smoothly with QSOs in the
Baldwin diagrams; when all luminosities are considered together, the
BE displays strong indications of curvature, although we note that
quantitative measures of this curvature will depend at some level on
the choice of cosmology ($q_0, H_0$) for calculation of quasar
luminosities.

The causes of curvature in the BE remain uncertain.  Wamsteker \&
Colina (1986) noted the similarity between curvature in the BE and a
similar curvature (i.e. nonlinear response; \S5.1) in the Baldwin
diagrams for individual variable sources.  They argued that both
phenomena could be interpreted as the result of a transition of the
BLR to a matter-bounded state for luminous sources.  While
matter-bounded nebular components may well contribute to curvature in
the Baldwin diagrams, a global transition to a matter-bounded state
generally predicts that the Ly$\alpha$/\c4 and C~{\sc iii}]/\c4 ratios should
{\sl decrease} at higher luminosities (Shields et al. 1995), in
conflict with the observed trends (\S3.2).  The relationship
between the ensemble BE and phenomenology of variable sources is
considered in more detail in the following section.  An alternative
explanation for curvature in the BE was advanced by Netzer et al.
(1992), who generated a similar pattern theoretically, using thin
accretion disks with random inclination and luminosity-dependent 
spectral energy distribution (SED).
As noted earlier (\S4.1), this model has been criticized on the basis
of its predicted distribution of $W_\lambda$ (Francis 1993); the
theoretical treatment of accretion disks (or other continuum sources)
in AGNs is also the subject of continuing discussion and debate.

\section{Variable Luminosity}

\subsection{The Intrinsic BE}

\begin{figure}
\vspace{3.4truein}
\includegraphics{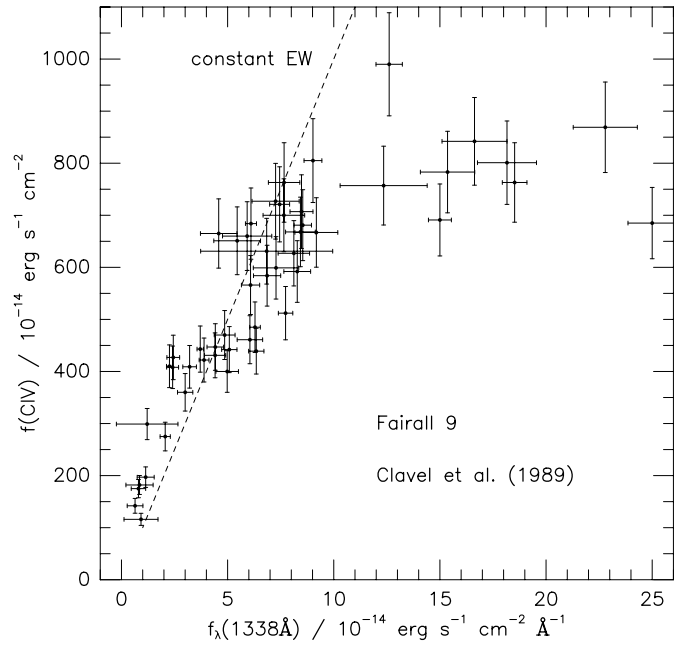}

\caption{Line flux for \c4 plotted as a function of 1338\AA\ continuum
flux for the variable Seyfert galaxy Fairall 9; measurements are taken
from Clavel et al. (1989).  Each point represents a different epoch of
observation.  No attempt has been made to remove time delay effects.
The dashed line corresponds to a constant $W_\lambda = 100$ \AA .}

\label{figA}
\end{figure}

Variability is now recognized as a probable source of scatter in the
ensemble BE, rather than its origin, and the observed (negative)
correlation between luminosity and equivalent width in {\sl individual}
variable sources has become known as the ``intrinsic
Baldwin Effect'' (Pogge \& Peterson 1992).  An intrinsic BE is
commonly observed for most of the strong UV lines in Seyfert galaxies.
Different lines exhibit different slopes, as in the ensemble case, and
there are some indications that the higher ionization lines again show
a systematically steeper relation.  For variable objects, the trend is
often expressed in terms of line versus continuum fluxes, rather than
equivalent width versus luminosity; the intrinsic BE then appears as a
nonlinear correlation in this plane (or a slope less than unity in a
log-log plot; see Figure~\ref{figA}).  The physics
underlying the intrinsic BE may be quite different from that of the
global trend, but the resemblance remains intriguing and the study of
variable sources offers additional relevant information on broad-line
region structure.

Several explanations have been advanced to account for the intrinsic
BE.  One possibility, discussed previously for the ensemble BE
(\S3.2), is that the trend arises from a luminosity dependence of
the ionizing continuum shape, such that the continuum becomes softer
when it brightens.  In this scenario, the optical/UV continuum
luminosity varies with greater amplitude than does the far-UV/X-ray
continuum; the line emission, responding to the high-energy radiation,
thus varies less strongly than the observable continuum, resulting in
a nonlinear relation between the two.

The luminosity dependence of continuum shape can be tested directly
for variable sources via simultaneous monitoring at UV and X-ray
energies.  Assembling the requisite data sets with reasonable temporal
coverage remains challenging, however, and at present only a handful
of sources have received such scrutiny.  The results are
heterogeneous: Fairall 9 exhibited some degree of correlation in the
expected sense, with variability amplitudes greater in the UV bandpass
than in X-rays (Clavel et al. 1989); but in NGC~5548 the fluxes at
1350 \AA\ and $2 - 10$ keV scaled together by similar factors (Clavel
et al. 1992), and an intensive monitoring campaign for NGC~4151 found
a {\sl higher} amplitude of variation at $1 - 2$ keV than in the
observable UV bandpass (Edelson et al. 1996).  The lack of a clear
signature is underscored by recent intensive monitoring of NGC 7469;
the UV and X-ray fluxes for this source varied with comparable
amplitude, and moreover exhibited only a weak degree of correlation
(Nandra et al. 1998).  Studies restricted to the optical/UV bandpass
have also examined the luminosity dependence of continuum shape, and
to the extent that any trend is present, the observable continuum
appears to get {\sl harder} with increasing luminosity in variable
sources (e.g., Edelson et al. 1990).

\begin{figure}
\vspace{4.8truein}
\includegraphics{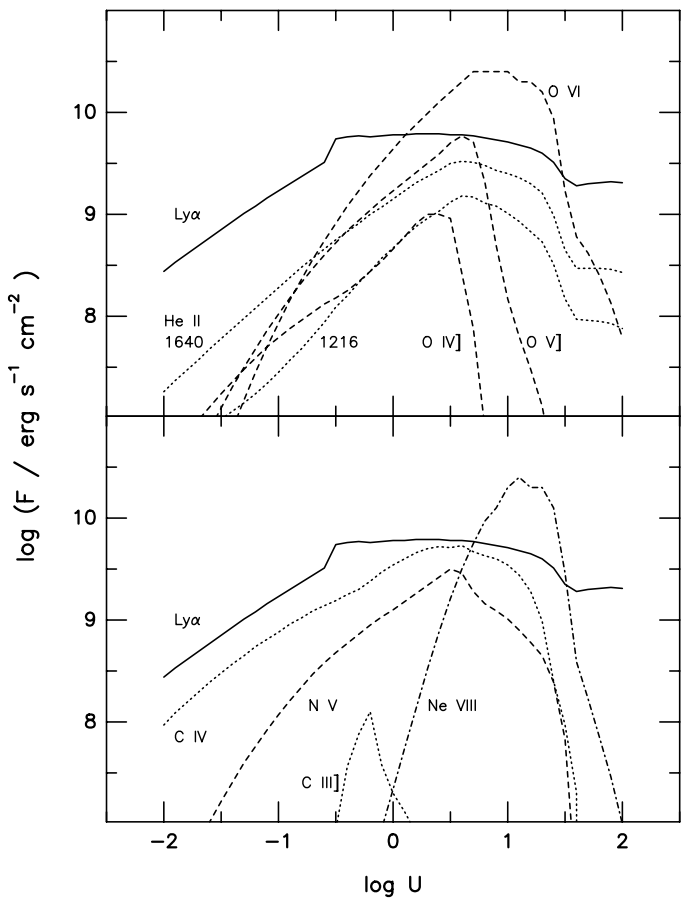}
\caption{Cloud emissivity in lines as a function of continuum
luminosity state.  The abscissa is expressed in terms of ionization
parameter $U$, the dimensionless ratio of ionizing photon and particle
densities at the irradiated cloud face.  The density of hydrogen (with
associated cosmic abundances) within the clouds is fixed at $10^{11}$
cm$^{-3}$, appropriate for the BLR, and column density $N = 10^{23}$
cm$^{-2}$.  The incident radiation field is described by a
representative AGN continuum shape (see Shields et al. 1995 for
details).  Slopes less than unity translate into a decrease in line
equivalent width $W_\lambda$ with increasing luminosity $L$ -- i.e.,
an intrinsic Baldwin Effect.}
\label{figB}
\end{figure}

A second potential cause of an intrinsic BE stems from ionization and
temperature effects in broad-line region clouds subject to a changing
radiation field.  The most straightforward example is provided by
matter-bounded (``optically thin'') clouds, which can undergo a global
change in characteristic ionization state as the ionizing continuum
fluctuates; this behavior contrasts with that of high column-density
clouds, which retain a region contributing intermediate- and
low-ionization emission for any continuum level.  For thin clouds, the
result can be a positive, null, or negative response in a given line
to changes in the continuum (Figure~\ref{figB}), with reduced response
contributing to an intrinsic BE.

Several lines of observational evidence support the presence of an
optically thin component in the BLR of AGNs.  These include: 
\begin{enumerate}

\item Evidence of negative line response for clouds in the inner BLR.
Sparke (1993) examined light curve data for NGC~5548, and noted that
the line autocorrelation function for prominent UV features was
narrower in width than the autocorrelation function for the continuum,
in conflict with expectations for photoionized clouds described by a
positive line response.  

\item The strength of high ionization emission
features such as O~{\sc vi} $\lambda$1034 and Ne~{\sc viii}
$\lambda$774.  The prominence of these features may require a
distinct, highly ionized, and presumably optically thin component
within the BLR (Netzer 1976; Davidson 1977; Hamann et al. 1998).  

\item The luminosity dependence of the \c4/Ly$\alpha$ line ratio.
Photoionization calculations for thick clouds generally predict that
this ratio will increase as the continuum brightens (e.g., Ferland \&
Persson 1989), while a number of Seyfert galaxies show the opposite
trend.  A very soft continuum can sometimes reproduce the observed
behavior (Gondhalekar 1992), but often has difficulty in
simultaneously matching the line equivalent widths.  A mix of thick
and thin clouds provides an alternative that can account for the
observed line responses and strengths (Shields et al. 1995).  

\item X-ray ``warm'' absorbers.  These highly ionized components are 
now known to be present in a large fraction of Seyfert nuclei (e.g.,
Reynolds 1997; George et al. 1998).  Variability and recombination
timescale arguments have been used in a few cases to bound the
absorber location to a scale comparable to the BLR (e.g., Otani et
al. 1996).  In these instances the absorber may represent the
high-ionization extension of the BLR; the same material would be
expected to emit efficiently in the Ne~{\sc viii} line.  Warm
absorbers are unambiguously matter-bounded systems.

\end{enumerate}

A further possible contributor to an intrinsic BE arises from radiative
transfer effects within the broad-line clouds.  Line photons with energy
sufficient to ionize hydrogen from the $n = 2$ state can be destroyed via
Balmer continuum absorption.  Balmer continuum opacity is expected to
scale in proportion to the ionizing flux incident on a cloud, so that
the efficiency of line destruction increases in higher luminosity states
(see Shields \& Ferland 1993 for details).  The result is a nonlinear
response in the line.  This effect is important for Lyman $\alpha$, as
demonstrated by explicit calculations, and is also expected to be relevant
to varying degrees for other lines.  At present, no means appear to
be available for isolating this behavior from other contributors to an
intrinsic Baldwin Effect.

A final basis for generating an intrinsic BE originates in light
travel-time effects in variable sources.  In the observer's frame,
line emission from circumnuclear clouds is expected to respond to
continuum variations with a delay, resulting from the added path length
for the continuum light to reach clouds outside our line of sight, plus 
path length variations for the emitted line radiation that depend
on the geometrical distribution of clouds.  Correlation analyses of
Seyfert light curves provide abundant evidence for such delays, which
form the basis for reverberation mapping studies of BLR structure.
Pogge \& Peterson (1992) have emphasized that the resulting phase
offset between the continuum and line light curves is a source of
scatter for the intrinsic BE; removal of a characteristic ``lag''
between the two light curves results in a tighter correlation between
line and continuum flux.

A complication to this general picture arises because there is
unlikely to be a single lag that is appropriate for all the emitting
components in the BLR.  The emission-line response to continuum
variations is both time-shifted and smeared by light travel-time
effects determined by the three-dimensional structure of the BLR.  As
a result, there will be a general tendency for some of the emission
components to be referenced to an inappropriate continuum level,
regardless of the choice of a lag.  The resulting phase offsets will
lead to larger $W_\lambda$ at low continuum states, and smaller
$W_\lambda$ in high states, compared to the predictions for idealized
clouds with no light travel-time delays.  The general role of such
delays in producing an intrinsic BE may be even stronger, however, if
significant emitting gas is present at large distances from the
continuum source, such that the light crossing time for the cloud
distribution is considerably greater than the timescale of
variability in the continuum.  In this case the emitting aggregate of
clouds may contribute a nearly constant line flux while the continuum
undergoes substantial variation, leading to a strong BE.  The
existence of substantial emitting gas at large radius within the BLR
can be difficult to exclude or constrain (e.g., Done \& Krolik 1996).

To summarize this section, several causes may contribute to the
intrinsic BE observed in variable AGNs.  These include a
luminosity-dependent continuum shape, the influence of matter-bounded
clouds, luminosity-dependent optical depth effects, and time delays
for light propagation across the BLR.  Light travel-time effects are
known to be operative at some level, while the importance of continuum
shape for the intrinsic BE is less certain.  The relevance of the
other two factors, thin clouds and optical depth effects, will depend
on the aggregate distribution of cloud properties in the BLR, since
reduced emissivity in one subset of clouds may be overwhelmed by
growth in emissivity elsewhere.  We consider this point in more detail
in the following section.

\subsection{The Intrinsic versus Ensemble BE}

Detailed comparisons of the global and intrinsic BEs have drawn
attention to the different slopes exhibited in Baldwin diagrams for
the two trends, with the intrinsic effect systematically steeper in
$W_\lambda$ versus $L$ (e.g., KRK).  As a quantitative example, AGN
ensembles plotted in the $W_\lambda$(\c4) versus continuum luminosity
plane typically display a logarithmic slope of $\sim -0.1$ to $-0.3$
(Korista et al. 1998 and references therein), while variable sources
display slopes of $\sim -0.3$ to $-0.9$ (KRK; Edelson et al. 1990).
While this contrast may be ascribed in general terms to the
differences between static and time-variable systems, some additional
examination of this issue is potentially informative.

Quasars and Seyfert galaxies have quite similar spectra, to first
order, which implies that their broad-line regions scale in some
nearly uniform sense, in terms of covering factor, velocity field,
density, ionization parameter, etc.  The ensemble Baldwin Effect tells
us that this scaling is not altogether homologous, although the
luminosity dependence is weak -- a factor of 10 variation in
$W_\lambda$(\c4) over 6 orders of magnitude in $L$.  Quasars evidently
undergo significant evolution in luminosity, and we can thus consider
the behavior of line emission as a Seyfert evolves to a quasar, or
vice versa.  A variable Seyfert nucleus (e.g., Fairall 9, with a
factor of 30 variation in luminosity; see Figure~\ref{figA}) arguably
represents this process in miniature.  That being the case, it is
perhaps surprising that the intrinsic and global BEs are so different,
i.e., Seyfert nuclei do not simply slide along the locus of the
ensemble relation as they vary.

\begin{figure}
\vspace{3truein}
\includegraphics{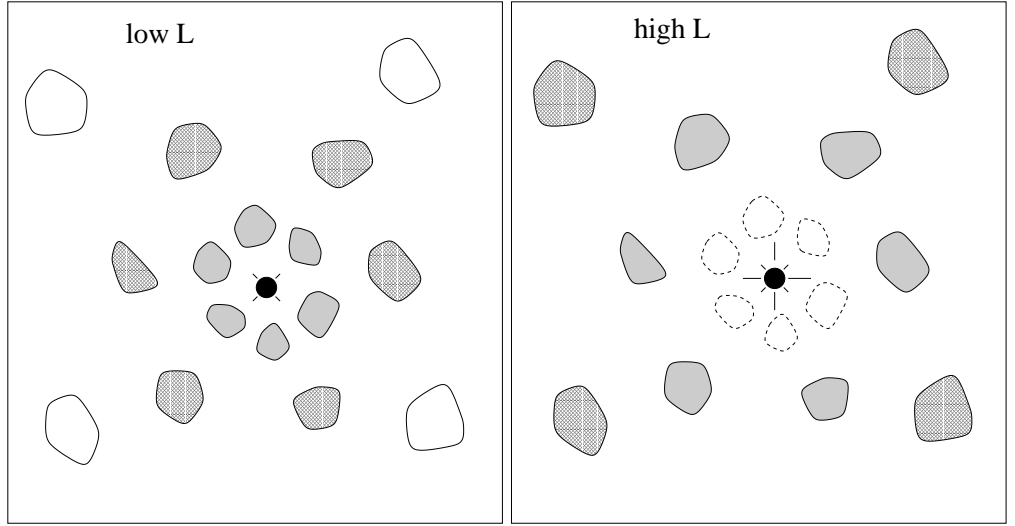}
\caption{Effects of covering factor versus radius, for variable luminosity.
If the lightly shaded clouds in this example are most efficient at emitting 
\c4, then $W_\lambda$(\c4) will drop as the continuum source brightens 
(producing a BE), due to the rapid falloff in fractional coverage.
}
\label{figC}
\end{figure}

Some perspective on the intrinsic versus ensemble behavior can be
derived from considering quasar luminosity evolution in conjunction
with the locally optimally-emitting cloud (LOC) model for the BLR
(Baldwin et al. 1995).  The essential idea of the LOC
model is that clouds emit with high efficiency in a given line only in
a rather restricted portion of the parameter space of cloud density $n$
and incident ionizing flux $\Phi$ (or distance $r$ from the continuum
source)\footnote{Column density represents another free parameter
describing the clouds, and the distribution of column densities will
influence the proportions of ionization- and matter-bounded clouds.}.
If we consider an ensemble of clouds surrounding a continuum source,
we can imagine correspondingly a zone in radius where clouds with the
appropriate density will dominate the total emission in, say, \c4; 
clouds at smaller radii are either too dense or too highly
ionized to emit strongly in this line, while clouds at larger radii
will have little C$^{+3}$ or lie outside the BLR.  If we were to
increase the luminosity of the central continuum source, the radius
describing the region of efficient \c4 emission would move
outward.  The effect on $W_\lambda$(\c4) would then depend on
{\sl cloud covering factor as a function of radius}\/ (see Figure~\ref{figC}).

If the covering factor $f_c$ of the clouds diminishes rapidly with
radius, we would expect the line equivalent width to decrease as the
source brightens (a Baldwin Effect); correspondingly, radial
distributions should also exist that would yield an {\sl increase} in
equivalent width for higher $L$ (an anti-Baldwin Effect).  What radial
distribution constitutes the intermediate case, corresponding to the
homologous BLR?  To obtain $W_\lambda$ independent of $L$, the
fractional coverage $\Delta f_c$ represented by clouds with a given
density, within an interval of incident flux $\Delta \Phi$, should be
independent of $L$.  The relevant differential covering factor is then
\begin{equation}
{{df_c}\over{d\Phi}} = {{df_c}\over{dr}}\,{{dr}\over{d\Phi}}~.
\end{equation}
Note that $r \propto \sqrt{L/\Phi}$, so that ${{dr}\over{d\Phi}} \propto
L^{1/2}\,\Phi^{-3/2}$.  If we express ${{df_c}\over{dr}} \propto 
r^\gamma$, then ${{df_c}\over{dr}} \propto L^{\gamma/2}\,\Phi^{-\gamma/2}$,
and
\begin{equation}
{{df_c}\over{d\Phi}} \propto {{L^{(1 + \gamma)/2}}\over{\Phi^{(3+\gamma)/2}}}~,
\end{equation}
which is homologous and independent of $L$ if $\gamma = -1$.

If the intrinsic Baldwin Effect ultimately stems from a steep fall-off
in circumnuclear covering factor ($\gamma < -1$), then the luminosity 
evolution of AGNs must be accompanied (perhaps unsurprisingly) by
substantial structural changes within the BLR, in terms of the distribution
of matter.  A less violent adjustment is required for a homologous profile,
which in turn provides a natural basis for producing grossly similar
Seyfert and quasar spectra.  The structure of $f_c(r)$ thus contains
potentially significant information on the physical evolution of AGNs.

An important independent constraint on $f_c(r)$ is available from
global fits to quasar spectra.  Baldwin (1997) has reviewed the LOC
model and its predictions for relative line strengths as a function of
$f_c(r)$, parametrized in terms of the differential power-law index
$\gamma$.  As can be seen from his Figure 1, good agreement with the
average quasar spectrum is obtained with $\gamma = -1$, which may
imply that the BLR is indeed homologous.  In this case the dominant
cause underlying the intrinsic BE is likely to be variability and
light travel-time effects, as discussed in the previous section.
Further comparisons of this type, perhaps including added constraints from
linewidth measurements, may provide stronger constraints on $\gamma$
and hence BLR structure, as well as the origins of the intrinsic BE.

\section{Further Systematics and Phenomenology}

\subsection{Line Profile Issues}

Another means of investigating the BE and its underlying physics is to
examine the luminosity dependence of emission as a function of
velocity within the line profile.  OPG examined difference spectra
between high- and low-luminosity composites, and demonstrated that
most of the variation in equivalent widths stems from changes in the
strength of the line core.  Another way of stating this result is that
the variable component has a full-width at half maximum (FWHM) that is
narrower than the overall profile.  This finding bears an intriguing
similarity to the results of the PCA analysis of FHFC, which found
that most of the variance in quasar line strengths (as quantified in
their first principal component) derived from modulations in the
emission line cores\footnote{A PCA analysis of {\sl IUE} spectra for
variable Seyfert nuclei revealed a quite different behavior, with the
line wings showing the strongest fluctuations (T\"urler \& Courvoisier
1998).  The contrasting behavior of the intrinsically variable sources
can probably be attributed to the strong influence of light
travel-time effects on the response across the line profile.}.  The
role of the core, or ``intermediate line region'' (ILR; Wills et al.
1993a) component in this regard is also suggested by significant
negative correlations in QSO spectra between line velocity width and
equivalent width (e.g., Francis et al. 1993; Wills et al. 1993a;
Brotherton et al. 1994b).  Detailed models of the ILR (Wills et al.
1993a; Brotherton et al. 1994a) predict only weak contributions to
the H$\beta$ and $\lambda$1400 features, consistent with the little
evidence for a BE in these lines (\S3.1).

Understanding the BE as purely a line-core phenomenon is not without
complications, however.  FHFC used their PCA eigenvectors to explore
the relative contributions of line cores and wings to the BE, and
found that the wings acted as an important contributor to the BE.
Additional evidence for a BE in the line wings has been presented by
Francis \& Koratkar (1995), who argued that the red wing of \c4 in
particular displayed a luminosity dependence (see also Corbin \&
Boroson 1996).  A common thread throughout these studies, however, is
that the strength of the BE (i.e. fractional change in $W_\lambda$) is
greater for the line core than for the line wings, in general accord
with the results from OPG.

If we take the link between line core variation and the BE at face
value, this finding may have important theoretical consequences.  If
the BE is ultimately a consequence of the luminosity dependence of the
continuum SED in QSOs (\S 3.2), it is perhaps surprising that the line
core and wings do not show comparable response to continuum hardness
variations.  This result may point to rather specific distributions of
cloud properties as a function of radius, if the velocity field of the
BLR is Keplerian (e.g., Brotherton et al. 1994a).  Korista et
al. (1997a) have published a grid of BLR $W_\lambda$ predictions as a
function of continuum SED and cloud properties, which might be
explored for this purpose; our brief inspection of their figures did
not lead to an obvious prescription for matching the observations,
however.  An alternative solution might entail anisotropic continuum
emission such that different cloud components within the BLR see a
different ionizing SED.  While this represents a more complicated
scenario, it would not be altogether surprising in light of
analyses implying that the ionizing continuum incident on the BLR
clouds is not the same as what we see (Korista et al. 1997b) -- which
opens up more general worries concerning the validity of invoking a
luminosity-dependent SED, derived from measurements, for explaining
the BE.

Finally, while the line core strength may be fundamental to defining
the BE, it is important to recognize that there is considerable
variation in the cores, and the overall line strengths, which is
apparently unrelated to source luminosity.  These variations thus
appear as scatter in the Baldwin diagrams; physically this range of
behavior may trace differences in orientation or structure of the BLR
and continuum source.  The extent to which core modulation and the BE
are in fact distinct phenomena is suggested by the properties of the
first eigenvector identified in the PCA analysis by FHFC.  In
comparison with the other eigenvectors, this component features
strong, relatively narrow emission lines, described by a large
Ly$\alpha$/\c4 ratio.  As a result, when this component weakens, the
line cores and overall equivalent widths diminish, which would be
consistent with the trend with increasing luminosity seen in the BE;
but the diminution of this component is accompanied by a {\sl
decrease} in the composite Ly$\alpha$/\c4 ratio, which runs counter to
the observed correlation with $L$ (\S3.2).  While the FHFC first
eigenvector may capture much of the variance in QSO spectra, other
parameters clearly come into play in establishing their
luminosity-dependent behavior.

\subsection{Radio-loud versus Radio-quiet}

The evidence suggesting that the strength of the BE is related to
source radio properties (\S2.2) is worth further consideration
in light of larger trends in quasar phenomenology.  Comparative
studies of radio-loud and radio-quiet QSOs have revealed a number of
detailed spectroscopic differences, and some of these disparities may
contribute to the contrast suggested in the Baldwin diagrams.  Several
analyses have established that radio-loud quasars display Ly$\alpha$
and \c4 lines with narrower profiles and larger average $W_\lambda$
than do radio-quiet sources (Francis et al. 1993; Wills et al. 1993a;
Brotherton et al. 1994b; Corbin \& Francis 1994).  Another way of
stating this result is that the emission features for the radio-loud
objects are more dominated by the line core than their radio-quiet
counterparts.  If the BE is driven primarily by changes in the line
core (\S6.1), then objects with emission lines dominated by the
core component -- i.e., the radio-loud objects -- might be expected to
show a cleaner correlation, consistent with the putative observational
trend.

\subsection{The X-ray BE}

One of the distinguishing features of quasars and AGNs is that they
emit over virtually the entire electromagnetic spectrum, from
$\gamma$-rays to radio wavelengths, and their optical/UV radiation is
only a fraction of the total emitted.  Furthermore, radiation at other
wavelengths almost certainly provides some of the most important keys
for understanding both the BE and the nature of quasars themselves.
Multi-wavelength observations are a vital part of quasar/AGN research.

The BE was discovered in the UV part of the spectrum, and
observational constraints as well as atomic physics have dictated that
studies of the luminosity dependence of broad-line emission are
conducted almost exclusively in the rest-frame UV/optical bandpass.
Improvements in the sensitivity and spectral resolution of X-ray
telescopes have resulted in the detection of an important new line
diagnostic of AGNs, in the form of the Fe K$\alpha$ line at 6.4 keV.  This
feature is believed to form by fluorescence in high column-density
material irradiated by an X-ray continuum.  The observed K$\alpha$
emission from Seyfert 1 nuclei can be very broad, and has been modeled
as fluorescence in a relativistic accretion disk (Nandra et al. 1997b
and references therein).  Narrow components are also observed, however
(e.g., Guainazzi et al. 1994), which may arise from structures such as
a circumnuclear torus of high column density matter (Krolik et
al. 1994; Ghisellini et al. 1994).

The recent availability of X-ray spectra of QSOs with decent spectral
resolution has made it possible to investigate the luminosity
dependence of Fe K$\alpha$.  Iwasawa \& Taniguchi
(1993) employed {\sl GINGA} spectra to argue that K$\alpha$ equivalent
widths were systematically weaker in more luminous systems, and drew
attention to the parallel with the ultraviolet BE.  While this finding
was subsequently challenged (Nandra \& Pounds 1994), a recent analysis
based on {\sl ASCA} spectra (Nandra et al. 1997a), reviewed by Paul
Nandra at this conference, lends strong support to a BE for the Fe
K$\alpha$ line.

Several interesting comparisons can be made between the detailed
phenomenology of the UV and X-ray BEs.  The existence of an X-ray BE
and its parallel in the UV lines initially led Iwasawa \& Taniguchi to
argue that this constituted evidence of a physical origin of the
K$\alpha$ emission in the BLR.  However, the X-ray correlation appears
to be strongly influenced by a luminosity dependence in the K$\alpha$
{\sl wings}, in contrast with the core-dominated UV trend.  Under the
existing interpretations of the K$\alpha$ profile, this behavior can
be taken as evidence that the X-ray BE is largely an accretion disk
phenomenon.  A connection between the K$\alpha$ emission and the BLR
may still be possible if the high-ionization broad lines are largely
produced in a disk structure (e.g., Murray \& Chiang 1998).  While
additional detailed physics is required to account for the different
line profile behaviors of the UV features and K$\alpha$ lines, a
further interesting similarity exists in that the BE for K$\alpha$ is
strongest in the red wing, in agreement with the UV trend.

Fluorescence of Fe K$\alpha$ in the AGN context is apparently part of
a larger pattern of spectral ``reflection'' signatures, resulting from
scattering and emission by X-ray-irradiated media with large column density,
which also include a Compton-reflection hump peaking in flux density
at $\sim 30 - 50$ keV and Fe K-edge absorption at $\sim 7$ keV; theoretical
studies predict an additional reflection component of thermal emission
that would contribute to the optical/UV ``Big Blue Bump'' (BBB;
Guilbert \& Rees 1988; Lightman \& White 1988).  Nandra et al. (1995)
have pointed out an important difficulty in associating the BBB with
reflection, in that luminous QSOs that typically exhibit a prominent
BBB (as parametrized, for example, by a steep $\alpha_{ox}$) also
exhibit {\sl weak} reflection signatures in the X-ray bandpass,
i.e. weak K$\alpha$ and Compton reflection features.  This result
strongly suggests that processes other than reflection/reprocessing of
the X-ray continuum dominate the generation of the optical/UV
continuum in AGNs.

The X-ray BE shows some dependence on source radio properties, with
radio-quiet QSOs featuring characteristically larger $W_\lambda$ for
K$\alpha$ than radio-loud sources show (Nandra et al. 1997a; Reeves et
al. 1997).  This result runs counter to the UV pattern.  A possible
means of reconciling these findings is to postulate a beamed component
to the X-ray continuum in radio-loud sources, which is seen by
us\footnote{This picture assumes that radio-loud objects for which our
line of sight falls outside the beam would generally be classified as
something other than quasars -- e.g., FR~{\sc ii} radio galaxies
(Barthel 1989).} and the broad-line clouds but not by the medium
responsible for Fe K$\alpha$ fluorescence.  This component would give
rise to greater heating of the BLR clouds and hence the observed
enhancement in UV line strengths, while strengthening the observed
X-ray continuum and thus diluting the K$\alpha$ equivalent widths.  An
anisotropic component to the ionizing continuum might well be expected
for radio-loud sources, which show evidence of relativistic outflows
that naturally produce beamed radiation fields in the observer's
frame.  An anisotropic component to the X-ray emission has been
suggested previously for other reasons (e.g., Browne \& Murphy 1987), and an
enhancement of this component in radio-loud sources would be
consistent with the fact that radio-loud quasars are
characteristically harder in $\alpha_{ox}$ than radio-quiet systems
(e.g., Zamorani et al. 1981).

AGNs show substantial variability in the X-ray continuum, and recent
work has also revealed detections of variability in Fe K$\alpha$
emission in individual sources (Iwasawa et al. 1996; Yaqoob et
al. 1996; Nandra et al. 1997c; see also Iwasawa \& Taniguchi 1993).
The existing studies do not provide a clear basis for drawing an
analogy between the behavior of K$\alpha$, and the intrinsic BE seen
in UV lines; the degree to which K$\alpha$ flux is correlated with the
continuum in variable sources is ambiguous at this point, and may be
a complicated function of velocity across the line profile.  Improved
measurements are of more general interest for study of AGN structure
on very small scales.  Iwasawa \& Taniguchi (1993) have also discussed
variability in relation to the global X-ray BE, and suggest that the
range in luminosity over which the correlation is observed ($\sim 4$
orders of magnitude) argues, as in the UV case, against variability 
as the source of the ensemble trend.

\section{The Baldwin Effect at High Redshift}

\subsection{Evidence of Evolution}

Given the difficulties and ambiguities in delineating the BE in QSOs
at low and moderate redshift, it is not surprising that only limited
attention has so far been given to the possibility that the BE evolves
with time.  Francis \& Koratkar (1995) combined {\sl IUE} data and optical
spectra of LBQS quasars to study the evolution of quasar spectra with
redshift over the interval from 0.4 to 2.2.  Both the low- and high-redshift
samples displayed a BE of comparable amplitude, as measured for \c4
and Ly$\alpha$.  However, these researchers found that the UV
spectra of radio-quiet quasars do evolve between redshift 2 and the
present in the sense that the high-redshift quasars have a population
with weak-line cores that is not seen in the {\sl IUE} sample.  The question
of evolution of spectral properties with redshift is important for at
least two reasons: 1) any evolution would affect conclusions drawn
from the BE about cosmological parameters, and 2) spectral evolution
could provide clues about the evolution of quasars and AGNs
themselves.  One difficulty in the interpretation of the findings by
Francis \& Koratkar rests with the {\sl IUE} sample, which was not
well-defined in any sense, but rather represented what was available
in the archive.  A similar analysis with better sample selection criteria
would be preferable.  This is a topic that definitely needs continued study.

\subsection{Preliminary Results at $z > 4$}

Quasars at the highest known redshifts potentially offer the strongest
leverage on cosmological parameters, and the opportunity for studying
high-$z$ QSOs has grown rapidly in recent years with the discovery of
an increasing number of sources at $z > 4$.  Observational results for
$z > 4$ QSOs were reviewed at this conference by Shields; in mid-1998
there are approximately 90 such objects reported in the published literature.
Spectroscopic properties for major subsets of these sources have been
reported previously by Schneider et al. (1991), Storrie-Lombardi et al.
(1996), and Shields \& Hamann (1997).

\begin{figure}
\vspace{3.3in}
\includegraphics{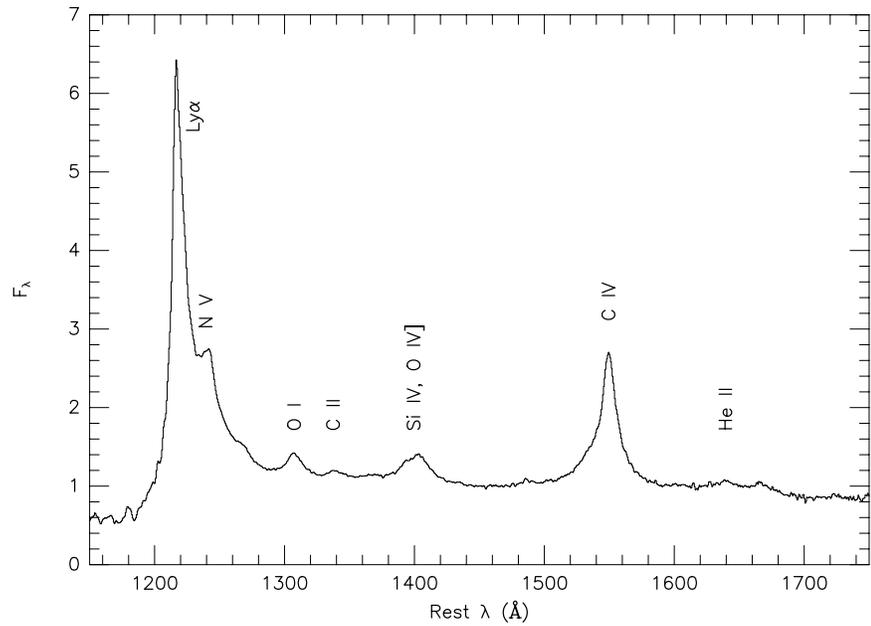}
\caption{Composite spectrum obtained by averaging spectra of 21 QSOs
at $z > 4$, observed with the Multiple Mirror Telescope (from Shields
et al. 1998).
}
\label{figOE}
\end{figure}

To first order, QSOs at high $z$ appear very similar spectroscopically
to sources at lower redshift.  A composite spectrum derived from
observations of 21 QSOs with $z > 4$ is shown in Figure~\ref{figOE}.
The apparent normality of the emission-line spectra of $z > 4$ quasars
is actually a striking feature when one considers that the lookback
time is greater than 90\% of the age of the universe (for $q_{0} =
0.5, \Lambda = 0$).  To the extent the emission spectra of
quasars reflect the abundances of elements like C, N, and O, one might
expect to see changes as a result of the chemical evolution expected in
the host galaxies of quasars over cosmological time scales.  In general,
however, order-of-magnitude differences in abundances lead to much smaller
variations in BLR line strengths in photoionization models, due to
thermostatic feedback effects.  An important exception may be present
in the \n5 $\lambda$1240 emission from QSOs, which may be sensitive to
secondary nitrogen enrichment (\S3.1), and does not display a BE.
The \n5 feature remains relatively strong at $z > 4$, pointing to
rapid early enrichment in the quasar nucleus environs (see Hamann
\& Ferland 1992, 1993 for details).  As discussed by Shields \& Hamann
(1997), these sources also show evidence of elevated O\,{\sc i}
$\lambda$1304 emission.  This feature forms primarily via fluorescence
pumped by H~Ly$\beta$ line coincidence, and is consequently expected to scale
with the O/H ratio, although factors other than metallicity
may also affect its strength.

\begin{figure}
\vspace{3in}
\includegraphics{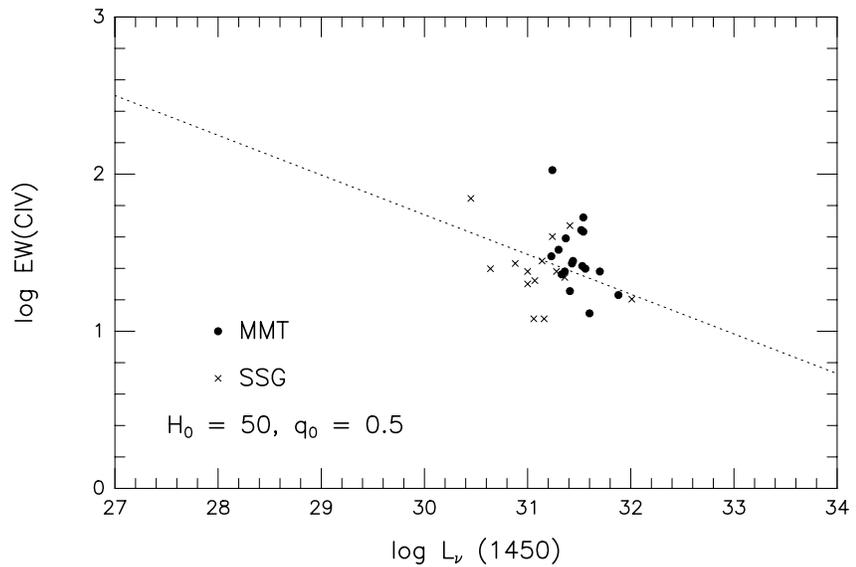}
\caption{ Baldwin diagram for quasars at $z > 4$.  The filled points
represent MMT observations from Shields et al. (1998), while the
$\times$'s represent data from Schneider et al. (1991).  The \c4
BE reported by OPG, based on measurements at lower redshift, is shown
by the dotted line.
}
\label{figOF}
\end{figure}

Quasars detected at $z > 4$ are almost invariably high-luminosity
objects, and provide useful data points at high $L$ for inclusion in
the Baldwin diagrams.  A summary of results to date is shown in
Figure~\ref{figOF} for the \c4 line, with data taken from Shields et
al. (1998) for a sample of objects that are largely selected on the
basis of optical colors (filled points), and from Schneider et
al. (1991) for grism-selected objects ($\times$).  The dotted line
represents the fit obtained to the BE by OPG for quasars at lower $z$.
Schneider et al. noted a tendency for many of their sources to fall
below the extrapolated BE in this diagram; with the addition of the
Shields et al. points, the results appear consistent with the
low-redshift fit, accompanied by the customary degree of scatter.

Selection effects remain a point of concern in comparing the $z >4$
findings with low-redshift results.  The color-selected objects in
particular are sensitive to inclusion of high-$W_\lambda$ objects that
are preferentially found in magnitude-limited surveys, due to the
added flux contributed by the emission line (typically Ly$\alpha$
observed in the $R$ bandpass\footnote{Note that the influence of the
Ly$\alpha$ line in this regard is exacerbated at large redshift by the
$(1 + z)$ scaling of observed $W_\lambda$, and the diminished
continuum blueward of the line caused by the Ly$\alpha$ forest.}; see
Kennefick et al. 1995 for quantitative details).  This bias may contribute
to a weak tendency for the color-selected sources to exhibit larger
equivalent widths than are found for the grism-selected objects in
Figure~\ref{figOF}\footnote{Interestingly, this pattern runs counter 
to the historical problems of selection bias, which favored large-$W_\lambda$
sources in grism-selected samples, in contrast with color selection
techniques (\S2.2).  Objective selection algorithms for
analysis of grism observations have largely removed this bias (e.g.,
Schmidt et al. 1995), while selection effects in high-redshift
color-selected samples may be nonnegligible.}.

The existing results for quasars at the highest known redshifts
nonetheless do not appear markedly different from their low-redshift
counterparts, in terms of the BE as seen in the strong UV lines.  We
note, however, that this statement is ultimately sensitive to the
choice of cosmology in computing $L$.  (Figure~\ref{figOF} assumes
$H_0 = 50$ km s$^{-1}$ and $q_0 = 0.5$.)  This ambiguity is inherent
in studying the BE and its possible evolution with samples in which
$L$ is strongly correlated with $z$; the resolution of these issues
awaits construction of samples extending over a wide range of $L$ at a
given $z$.  Researchers can also aid their colleagues by publishing
{\sl actual measurements of continuum flux and redshift}, rather than
simply the derived quantity $L$, which depends on the specific choice
of cosmology as well as measured $z$.

\section{Summary and Future Directions}

The Baldwin Effect is real and has important implications for the
emission-line regions of quasars.  Additional work will be needed if
we are to fully extract the meaning of this correlation.  The
existence of an ionization dependence in the BE provides important
clues to the underlying physics.  Variations in the BE across line
profiles, possible differences between radio-loud and radio-quiet BEs,
and constraints resulting from the intrinsic BE will also play a role
in sharpening our understanding of this phenomenon.  A clearer
understanding of the physical basis for the BE would gratify quasar
enthusiasts while strengthening the credibility of using this
phenomenon for calibrating cosmological diagnostics.

Several avenues of inquiry would be of particular value for making
progress in our understanding.  Large data sets with well-defined
selection criteria and uniform follow-up would aid in refining BE
measurements and their relation to radio properties and the ionizing
SED.  The Sloan Digital Sky Survey and surveys with the new generation
of X-ray telescopes ({\sl AXAF}, {\sl XMM}) can be expected to deliver
appropriate samples, while the burgeoning number of 8-m class telescopes 
will in principle make it possible to obtain follow-up spectroscopy
with the high signal-to-noise ratio necessary for reliable and reproducible
line measurements.  X-ray studies will also be important for probing
the X-ray BE with well-defined samples, while exploring the connection
between K$\alpha$ behavior and the UV lines.  

Beyond simply searching for the BE, it is now time to vigorously
pursue the origins of dispersion in the correlation.  The substantial
scatter has raised large questions about the suitability of the BE as a
cosmological tool.  Multiparameter and principal-components
investigations offer powerful tools for extracting the full details of
correlated phenomenology, which in turn can serve as inspiration and
testbeds for the construction of detailed physical interpretations.
The future promise of the Baldwin Effect as an entree to understanding
quasars and cosmology appears bright.

\acknowledgements

We thank our hosts for organizing the meeting.  Their hospitality was
in the best Chilean tradition.  Our colleagues provided valuable
commentary and lively discussion, which helped us all improve our
understanding of this important but complex topic in quasar research.
We wish to also thank Brad Peterson and Mike Brotherton for providing
helpful comments on a draft of this manuscript.  Figures 1, 2, and 4
are reproduced with the generous permission of the original authors,
Jack Baldwin, Paul Francis, \& Anne Kinney, respectively.  Support for
our research is provided by the NSF to PSO through grant AST-9529324,
and by NASA to JCS through grants NAG-3563 and NAG-3690.

\end{document}